# Universal Constants, Standard Models and Fundamental Metrology

G. Cohen-Tannoudji[1]

[1]Laboratoire de Recherche sur les Sciences de la Matière (LARSIM) CEA-Saclay, F91191 Gif sur Yvette Cedex

**Abstract.** Taking into account four universal constants, namely the Planck's constant $h$, the velocity of light $c$, the constant of gravitation $G$ and the Boltzmann's constant $k$ leads to structuring theoretical physics in terms of three theories each taking into account a pair of constants: the quantum theory of fields ($h$ and $c$), the general theory of relativity ($c$ and $G$) and quantum statistics ($h$ and $k$). These three theories are not yet unified but, together, they underlie the standard models that allow a satisfactory phenomenological description of all experimental or observational data, in particle physics and in cosmology and they provide, through the modern interpretation of quantum physics, fundamental metrology with a reliable theoretical basis.

## 1 Introduction

The scientific revolution that arose at the beginning of the 20$^{\text{th}}$ century coincided with the discovery or the re-discovery of universal constants that were interpreted as reflecting some fundamental limitation principles. It appeared that these limitations are not to be considered as insuperable obstacle but rather as *horizons* (1), which means that there exists some rationale to theoretically think what reality lies behind or beyond them. Actually, as says Ferdinand Gonseth (1890-1975), a Swiss mathematician philosopher, it seems that the whole of human knowledge is bounded by such horizons (2): "The previous results have a value that goes beyond the framework of geometry. They concern the entirety of knowledge, we mean the state in which knowledge comes to us, at a given instant: Nothing authorizes us to think that our knowledge, even at its last frontiers, is more than a knowledge horizon; that the last 'realities' that we have conceived are more than a reality horizon." In this chapter, we intend to show how this concept of reality horizon, related to universal constants allows establishing a strong relation between fundamental physics and metrology: just as a sailor is able to determine his or her position on earth by measuring the height of the sun with respect to the horizon, structuring theoretical physics in terms of reality horizons determined by dimensioned universal constants allows establishing a *comprehensive reliable frame of reference* consisting of *standard models* that depend on a finite number of dimensionless universal constants, the determination of which is the task of fundamental metrology.

## 2.1 Universal constants and the role of fundamental metrology

### 2.1 The interpretation of universal constants may change according to the evolution of theories

What one calls fundamental constants are the fixed and universal values of some physical observables (dimensioned or not) that enter the theories designed to describe the physical reality. There are several possible interpretations of these constants. They can appear as parameters in the framework of some models. They can lead to the unification of previously independent domains. They can serve as conversion factors between physical units. They can also, and this is the main role they can play in structuring physics, reflect some fundamental limitation principles. Actually, the interpretation changes according to the evolution of theories. This is the case for the four constants we shall consider in this chapter.

2.1.1 The vacuum velocity of light $c$

In Maxwell's electromagnetic theory of light, *c* appeared as a phenomenological parameter, with dimension of a velocity, related to the electric and magnetic properties of the "vacuum". It is when Hertz reformulated the Maxwell's theory in terms of the propagation of electromagnetic waves, and when he experimentally discovered that the propagation velocity of these waves was compatible with the already measured velocity of light, that *c* got the status of a universal constant equal to the velocity of light in the vacuum, leading to the unification of electricity, magnetism and optics.

Later on, in 1905, *c* got a new interpretation, the one of an upper bound on any propagation velocity, translating a *fundamental limitation principle*, namely the *absence of instantaneous action at a distance*. The theory of special relativity takes this fundamental limitation into account by means of the concept of space-time, introduced in 1908 by Minkowski, and once this theory has been established, *c* got yet another interpretation, namely the one of a conversion factor between length and time units.

2.1.2 The Newton's constant *G*

The constant *G* allows unifying celestial and terrestrial mechanics in the Newton's theory of universal gravitation. In general relativity, the Einstein's theory of universal gravitation that encompasses the Newton's one and gives it back as an approximation in the low field limit, *G* is a coupling constant relating the matter energy-momentum tensor and the Einstein's curvature tensor. It may also be interpreted as a constant relating the mass of a black hole to the radius of its event horizon. When associated with *c* and with the Planck's constant *h*, it leads to the definition of the Plank's units of mass, space and time, characteristic of fundamental limitation principles related to quantum *gravity*

2.1.3 The Boltzmann's constant *k*

According to the current interpretive consensus [3], the Boltzmann's constant *k* is just a conversion factor relating temperature and energy: at thermal equilibrium, temperature is proportional to the average kinetic energy of molecules, and the proportionality factor is *k*. Since temperature is not considered as a fundamental physical quantity, *k* is not considered as a fundamental universal constant, it is a conversion factor allowing us to express temperatures in energy units. The interpretation of the universal constants that we are going to propose is at odds with this interpretive consensus, in particular about the status of the Boltzmann's constant. Actually it turns out that the history of the discovery and the interpretation of this constant made to appear several interpretations that do not reduce to the one of it being a conversion factor. Consider first the interpretation by Boltzmann of entropy in the framework of statistical thermodynamics: this physical quantity expresses the lack of knowledge we have about a system obeying the deterministic laws of rational mechanics but in which there exists a large number of microscopic configurations (called complexions) leading to the same macroscopic state; entropy *S* is proportional to the logarithm of the number *W* of complexions, and *k* (which has thus the dimensional content of entropy) is the proportionality factor,

$$S = k \operatorname{Ln} W \qquad (1)$$

This equation that is the fundamental equation of statistical thermodynamics has been extensively used by Einstein in the beginning of 20$^{th}$ century [4] for the purpose of two applications: one in an attempt to estimate the fluctuations affecting a system at thermal equilibrium in order to have an insight on the fundamental properties of the microscopic constituents of matter (atoms or molecules) [5], and another one in the interpretation of the Planck's formula on the black body radiation in terms of energy quanta [6].

In the framework of information theory, developed in the middle of the 20$^{th}$ century, the Boltzmann's constant got yet another interpretation, the one of the minimal cost, expressed in entropy of a single bit of information. If information is considered as a fundamental physical quantity, the Boltzmann's constant thus translates a fundamental limitation principle.

2.1.4 The Planck's constant *h*

It is precisely when trying to use the fundamental equation of statistical thermodynamics (1) to explain the spectrum of the black body radiation that Planck was led to introduce the constant now known as the Plank's constant, the elementary quantum of action $h$. The formula he derived realizes, by means of the three universal constants $c$, $h$, and $k$ (that Planck named as the Boltzmann's constant), the unification of statistical thermodynamics and electromagnetism. In the Planck's formula, as well as in its interpretation in terms of energy quanta proposed by Einstein in 1905, the Planck's constant appears as a proportionality factor between energy and frequency. According to the particle/wave duality, the Planck's constant is interpreted as a conversion factor between particle kinematics (energy, momentum) and wave kinematics (frequency, wave vector). More generally, in quantum physics, the Planck's constant translates a fundamental limitation principle stating that conjugate variables, i.e. the product of which has the dimensional content of an action, cannot be determined with arbitrary accuracies, but that these accuracies are constrained by the Heisenberg's inequalities.

## 2.2 A purely theoretical interpretation: the cube of theories

2.2.1 Two classes of universal constants

According to the program of rational mechanics, there are three, and only three fundamental physical units entering the dimensional content of all physical quantities, namely the mass M, the length L and the time T units. Already in 1899, when he had an idea about the constant that bears his name, Planck remarked that by means of combinations of the three dimensioned constants $c$, $G$ and $h$, one can build three fundamental dimensioned quantities, now known as the Planck's scales of length, time and mass, namely

$$l_P = \left(\frac{hG}{c^3}\right)^{1/2} \approx 10^{-35} m$$

$$t_P = \left(\frac{hG}{c^5}\right)^{1/2} \approx 10^{-43} s \qquad (2)$$

$$m_P = \left(\frac{hc}{G}\right)^{1/2} \approx 10^{19} GeV/c^2$$

and, by means of the Boltzmann's constant he also introduced a "Planck's scale of temperature", $T_P = m_P c^2 / k$. Once these fundamental units are fixed, all other universal constants can be considered as dimensionless: the mass of the electron, say, is some dimensionless fraction of the Planck's mass. Now, since the dimensioned universal constants seem to be related to fundamental limitation principles, it seems natural to associate them to some theoretical or axiomatic framework, and to associate the dimensionless constants with parameters in the framework of models. As long as these models are heuristic or phenomenological models, the constants are considered as adjustable parameters, but when a model, compatible with the axiomatic framework agrees well with all available observational or experimental data in a given phenomenological realm, the model acquires the status of a "standard model" and the adjustable parameters on which it depends are fixed and acquire the status of universal constants, which, hopefully would be calculable in the framework of an ultimate theory able to account for all dimensioned constants.

2.2.2 The "cube of theories"

According to the current interpretive consensus, there are, just as according to the program of rational mechanics, three and only three fundamental units, and thus three and only three dimensioned universal constants, $h$, $c$ and $G$. Entropy is considered as a dimensionless quantity; the Boltzmann's constant is considered as a conversion factor allowing to express temperatures in energy units; statistical physics is not considered as fundamental. The theoretical framework is structured as shown in fig.1, according to the way how the three fundamental constants are taken into account (i.e. set to 1) or not (i.e. set to zero). At the origin of the frame, the three constants $h$, $G$ and $1/c$ are set to zero, which represents *non gravitational Newtonian mechanics*; we have then the three theories that account for the three constants one by one separately, namely *special relativity* in which $1/c$ is set to 1, whereas $h$ and $G$ are set to zero, *Newtonian gravity* in which $G$ is set to 1 whereas $1/c$ and $h$ are set to zero and *quantum mechan*ics in which $h$ is set to 1 whereas $1/c$ and $G$ are set to zero; we have then three theories taking into account pairs of universal constants, namely the *quantum theory of field*s ($h$ and $1/c$ set to 1, $G$ set to zero), *general relativity* ($1/c$ and $G$ set to 1, $h$ set to zero), and a theory which does not seem very meaningful from the interpretive point of view, namely *non relativistic quantum gravity* ($h$ and $G$ set to 1, $1/c$ set to zero); finally the summit of the cube, opposite to the origin, represents what is considered as the ultimate goal of theoretical physics, a theory of *quantum relativistic gravity* which would take into account the three constants and which would hopefully allow determining the values of all dimensionless constants.

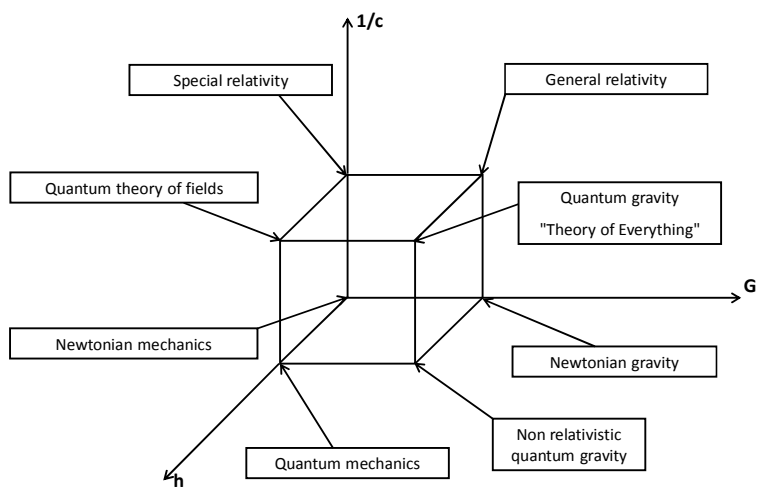

**Fig. 1.** The "cube of theories"

## 2.3 An interpretation of universal constants relevant to fundamental metrology

Obviously this purely theoretical interpretation is not suitable to discuss the role of fundamental metrology. On the one hand, it suggests that the three fundamental units are the Planck units which are far from being "natural" (for instance, a particle with the Planck energy would have the kinetic energy of a transportation aircraft), and a theory able to take into account the three universal constants $h$, $c$ and $G$ is not yet at hand. On the other hand, the link between the theoretical framework and phenomenology is not taken into account in this interpretation. In our opinion, this lack is due to an underestimation of the role of information as a genuine physical quantity precisely providing this link between theory and phenomenology. According to its informational interpretation, entropy is the information that is sacrificed when statistical averaging or *coarse graining* is performed. Now it turns out that in the *modern interpretation of quantum physics*, coarse graining is unavoidable in order for quantum mechanics to provide probabilities for different sets of alternative decoherent histories: decoherence may fail if the graining is too fine; moreover a coarser graining is necessary for having approximate classical predictability. If thus one wants to have an interpretation of universal constants that is consistent with this modern interpretation of quantum physics, that is so basic in fundamental metrology, one has better to include among the fundamental physical quantities the one that depends on coarse graining, that is entropy or information, and thus to include among the universal constants that determine the relevant theoretical framework, the Boltzmann's constant that has the dimension of an entropy.

2.3.1 The *c*, *h*, *k* scheme: kinematics versus dynamics.

Such an interpretive scheme, suitable for fundamental metrology is the one adopted by Christian Borde [7] who distinguishes a *kinematical* framework, framed by the three dimensioned constants *c*, *h* and *k* interpreted as reflecting fundamental limitation principles from a *dynamical* framework, in which interactions are described by means of dimensionless constants. Relativity allows the value of *c* to be fixed and the standard of length to be redefined, quantum mechanics allows the value of *h* to be fixed and the mass standard to be defined and statistical mechanics allows the value of *k* to be fixed and the scale of temperature to be defined. Interactions are then implemented in the dynamical framework and all the "quantities related to the various forces of nature can be formulated from the mechanical units and coupling constants without dimensions".

2.3.2. The (*h*, *c*), (*G*, *c*), (*h*, *k*) scheme: fundamental versus emergent physics.

The interpretation we propose in terms of a tripod of theories each taking into account a pair of universal constants is compatible with the scheme just described; it enlarges it in a way that allows to emphasize the role of modern quantum physics and fundamental metrology in the establishment of the standard models of particle physics and cosmology, and in the search of any new physics not described by them, in the hope of eventually unifying general relativity and quantum theory.

This interpretation is schematized in fig. 2: quantum field theory (QFT taking into account *h* and *c*) that generalizes quantum mechanics and underlies the standard model of particle physics and general relativity (GR taking onto account *G* and *c*) that generalizes relativity and underlies the standard model of cosmology are not yet unified but, together, they constitute the current conceptual framework of *fundamental physics*. Quantum statistics (QS taking into account *h* and *k*) that generalizes statistical mechanics and underlies modern quantum physics rather provides the conceptual framework of *emergent physics*. Figure 2 is meant to show that *fundamental metrology evolves at the intersection of fundamental and emergent physics*. This is what we intend to explain in more details in the following sections of this chapter.

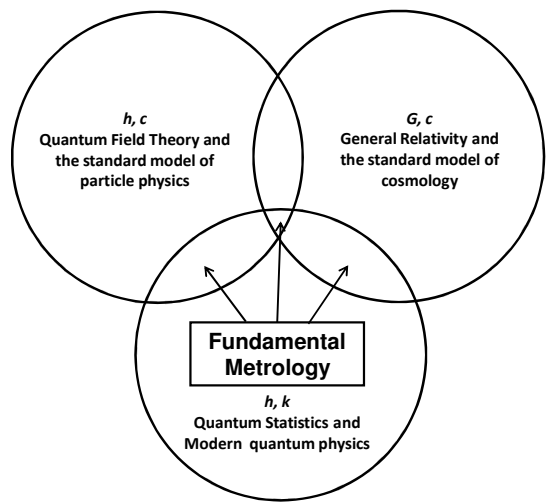

Fig.2 Fundamental metrology at the intersection of fundamental and emergent physics

## 3 Quantum Field Theory and the Standard Model of Particle Physics

### 3.1 What is Quantum Field Theory?

3.1.1 Fundamental implications of QFT

Quantum Field Theory (QFT) is the result of the merging of quantum mechanics and special relativity. In ordinary quantum mechanics (QM), the space coordinates of a particle are operators, whereas time is a continuous parameter. QM is local in time but not local in space. In QFT, the four space-time coordinates are continuous parameters, just as in classical relativistic field theory. QFT can be made local in space-time.

The relativistic generalizations of the Schrödinger equation, the Dirac equation $(i\gamma^\mu \partial_\mu - m)\psi(x) = 0$ or the Klein Gordon equation $(\Box - m^2)\psi(x) = 0$, admit negative energy solutions difficult to interpret if $\psi(x)$ is the wave function of an isolated particle. This problem is the first one met when one has tried to reconcile quantum mechanics and relativity. It is solved if Dirac or Klein Gordon equations apply to *quantum fields* rather than to wave functions of isolated particles. A quantum field is a field of *operators*, creating or annihilating, at any point of space-time a particle or an *antiparticle*. A particle and its antiparticle have the same mass, the same spin but opposite charges. *A particle with negative energy going backward in time is replaced by its antiparticle, with positive energy going forward in time*.

Particles and antiparticles are not material points: in fundamental interactions, the number of particles is not conserved; particles or antiparticles can appear or disappear. The creation or the annihilation of a particle or an antiparticle is a *quantum event* that causality and relativity force us to treat as *strictly localized* in space-time. Heisenberg indeterminacy inequalities imply that such strictly localized events *cannot be individually predicted in a deterministic way*. The predictability in QFT concerns *statistical ensembles of events*.

In the framework of QFT two fundamental theorems have been established:
- The spin/statistics connection theorem that states that fermions have a half integer spin, and that bosons have an integer spin
- The *PCT* theorem that states that all interactions are invariant under the *PCT* symmetry that is the product of *space parity P* by *time reversal T* and by *charge conjugation C*

3.1.2 The Fock space

In absence of interactions, (i.e. for free fields) the equations of motion (Dirac, Klein Gordon or Maxwell equations) can be solved entirely. In momentum space, the free field Hamiltonian reduces to an infinite sum of independent harmonic oscillators (normal modes). Quantum statistics (Bose-Einstein or Fermi-Dirac) are most easily taken into account by means of the Fock space, namely a Hilbert space in which the quantum states of the field are defined in terms of the *occupation numbers*, i.e. the number of field quanta with a given four momentum.

### 3.2 The Path Integral formulation of QFT

In a 1948 paper, untitled *Space-time approach to Non Relativistic Quantum Mechanics*, Feynman [8] re-formulates and re-interprets quantum mechanics as a *field* theory in terms of an integral over all the trajectories that can virtually follow a particle (the so-called Path Integral formulation):
- Thanks to the principle of *superposition of probability amplitudes*, the wave function is interpreted as a *field amplitude* that is evaluated by means of the Huygens principle
- It is the *field intensity* (proportional to the modulus squared of the amplitude) that is given a probabilistic interpretation
- This reformulation can be extended to the quantization of relativistic field theories in a way that preserves Lorentz invariance.

In QFT, the path integral (PI) is a *functional integral,* i.e. an integral over the fields of the exponential of *i* times the "classical" action in units of $\hbar = h/2\pi$. The "classical" action integral is the invariant integral over space-time of the Lagrangian, expressed in terms of c-number fields, which would lead through the least action principle to the "classical" equations of motion. In absence of interactions one recovers with PI the results of the standard quantization method for Klein Gordon, Dirac and Maxwell fields. When the Lagrangian involves an interaction term consisting of the product of fields multiplied by a small coupling constant, PI reduces to an expansion, called the *perturbative expansion*, in powers of this small coupling constant, the coefficients of which involve only ordinary integrals.

## 3.3 The paradigm of Quantum Electrodynamics (QED)

3.3.1 The QED Lagrangian

QED is the quantum relativistic theory of the electromagnetic interaction of the electron field, $\psi(x)$.

From the Dirac Lagrangian for this electron field

$$\mathcal{L}_{\text{Dirac}} = \bar{\psi}(x)\left(i\gamma^\mu \partial_\mu - m\right)\psi(x) \tag{3}$$

we derive the Dirac equation and a possible conserved current:

$$\begin{aligned}(i\gamma.\partial - m)\psi &\equiv (i\slashed{\partial} - m)\psi = 0; \bar{\psi}\left(i\gamma.\bar{\partial} + m\right) = 0 \\ \Rightarrow \partial_\mu j^\mu &= 0 \text{ if } j^\mu = -e\bar{\psi}(x)\gamma^\mu \psi(x).\end{aligned} \tag{4}$$

The QED Lagrangian is obtained by adding to the Dirac Lagrangian the Maxwell Lagrangian and an interaction Lagrangian:

$$\begin{aligned}\mathcal{L}_{\text{QED}} &= \mathcal{L}_{\text{Maxwell}} + \mathcal{L}_{\text{Dirac}} + \mathcal{L}_{\text{Interaction}} \\ \mathcal{L}_{\text{Maxwell}} &= -\frac{1}{4}F_{\mu\nu}F^{\mu\nu}; F_{\mu\nu} = \partial_\mu A_\nu - \partial_\nu A_\mu \\ \mathcal{L}_{\text{Dirac}} &= \bar{\psi}\left(i\gamma^\mu \partial_\mu - m\right)\psi \\ \mathcal{L}_{\text{Interaction}} &= -e\bar{\psi}\gamma^\mu A_\mu \psi.\end{aligned} \tag{5}$$

3.3.2 The perturbative expansion

The perturbative expansion is obtained through the *generating functional*

$$\mathcal{Z}\left(j_A, j_\psi, j_{\bar{\psi}}\right) = \int \mathcal{D}\psi \mathcal{D}\bar{\psi}\mathcal{D}A \exp\left\{\frac{i}{\hbar}\int d^4 x \left(\mathcal{L}_{\text{QED}} + j_\psi \psi + j_{\bar{\psi}}\bar{\psi} + j_A A\right)\right\},$$

where the *j*'s are non dynamical sources of the fields, and the integration symbol $\mathcal{D}$ represents a functional integration over field configurations.
- The power expansion in the sources generates the *Green's functions*.
- The power expansion of the Green's functions in power of *e* is the perturbative expansion the coefficients of which are sums of Feynman amplitudes associated with Feynman diagrams.
- The power expansion in $\hbar$ leads to the calculation of the effects of quantum fluctuations.

3.3.3 Renormalisability and renormalisation

Due to the occurrence, as implied by micro-causality, of products of fields evaluated at the same event (space-time point), the integral necessary to define the Feynman amplitudes associated with Feynman's diagrams containing loops, in general diverge, which could endanger the whole program of

QFT. Apart from the difficulty of negative energy that we discussed above, the problem of infinities in the perturbative expansion has been considered as the major drawback of QFT. The way out of this difficulty relies on the *renormalisation procedure* that we sketch very briefly here.

In QED, this procedure consists in three steps
a. Regularizing the divergent integrals by means of an energy "cut-off" to discard virtual processes involving arbitrary high energy
b. Splitting the values of the parameters entering the Lagrangian, the mass $m$ and the charge $e$ of the electron into their *bare* values, $m_0$ and $e_0$, i.e. the values they would have in absence of interaction, and their *renormalized* values, $m_R$ and $e_R$, i.e. the values they acquire due to the interaction.
c. It may happen that when one tries to express the Green's functions in terms of the (unphysical) bare parameters, the cut-off parameter cannot be sent to infinity without getting an infinite answer whereas, when expressed in terms of the (physical) renormalized parameters, the Green's function go to a finite limit when the cut-off goes to infinity. A theory in which such a circumstance occurs for all Green's functions and at all orders of the perturbative expansion is called *renormalisable*, and as such it is free from the problem of infinities.

It has been proven that QED is renormalisable, which means that this theory has a predictive power such that it can underlie a genuine standard model: the renormalized parameters cannot be predicted, but once they are determined from experiment, there are physical quantities that can be measured experimentally and theoretically predicted. In QED, the agreement between experiment and theoretical predictions is surprisingly good.

However there remains a problem: the renormalized parameters depend on an arbitrary energy scale related to the coarse graining implied by the experimental conditions, whereas one expects the predictions of a "fundamental" theory not to depend on the coarse graining. Fortunately, in a renormalisable theory, the *renormalization group equations* constrain the dependence of the renormalized parameters on the coarse graining in such a way that the physical predictions of the theory actually do not depend on it.

3.3.4 Gauge invariance in QED

The success of QED was considered as a miracle that it was very tempting to generalize to the theory of other fundamental interactions. It was then necessary to uncover the property of QED that could be at the origin of the miracle of renormalisability and that could be generalized to other interactions. Gauge invariance is precisely this property. Let us first remind how in classical electrodynamics, gauge invariance is related to current conservation. We write

$$\mathcal{L}_{\text{Classical}} = \mathcal{L}_{\text{Maxwell}} + ij_\mu A^\mu$$
$$\mathcal{L}_{\text{Maxwell}} = -\frac{1}{4} F_{\mu\nu} F^{\mu\nu} \;;\; F_{\mu\nu} = \partial_\mu A_\nu - \partial_\nu A_\mu, \tag{6}$$

This is invariant under the *gauge transformation* that consists on adding to the potential (that we now call the gauge field of the electromagnetic interaction) the four-divergence of an arbitrary function: $A_\mu \to A_\mu + \partial_\mu \lambda(x)$, if and only if the current is conserved. Consider now the symmetry of the Dirac Lagrangian $\mathcal{L}_{\text{Dirac}} = \bar{\psi}(i\gamma^\mu \partial_\mu - m)\psi$, it is invariant under the change of the phase of the electron field

$$\psi(x) \to \exp(-i\alpha)\psi(x)$$
$$\bar{\psi}(x) \to \bar{\psi}(x)\exp(i\alpha). \tag{7}$$

Because of the space-time derivative, the phase invariance is global and not local, which means that the phase changing factor cannot depend on $x$.

The interaction term may allow to impose a *local phase invariance* by compensating (infinitesimally) the local change of phase induced by the derivative by a local gauge transformation of the potential; in fact the QED Lagrangian is invariant under the following transformation, which one now calls the *local gauge invariance of QED*

$$A_\mu \rightarrow A_\mu + \partial_\mu \lambda(x)$$
$$\psi(x) \rightarrow \exp(-i\alpha(x))\psi(x) \tag{8}$$
$$\bar{\psi}(x) \rightarrow \bar{\psi}(x)\exp(i\alpha(x))$$
$$\text{iff } \alpha(x) = \lambda(x)$$

The very important feature is that one could have done the reasoning the other way around: one would have started from the Dirac Lagrangian for a spinorial field; one would have noticed its global phase invariance; one would have imposed a local phase invariance through the coupling with a gauge field to compensate the variation induced by space-time derivative; one would have completed the Lagrangian by adding the Maxwell Lagrangian describing the propagation of the gauge field thus introduced. *One would have thus recovered exactly the QED Lagrangian!*

Due to the fact that local gauge invariance is crucial in the proof of renormalisability of QED, it is very tempting to make of this invariance a unifying principle, a "road map" towards tentative renormalisable theories of other fundamental interactions:
- For a given fundamental interaction, identify the matter fields and their global symmetries,
- make these symmetries local through the coupling with suitable gauge fields,
- complete the Lagrangian with the propagation terms for these gauge fields,
- verify the renormalisability of the thus derived theory.

**3.4 The gauge theories of the standard model**

The construction of the standard model of particle physics has followed essentially this road map. Yang and Mills [9] first generalized the local gauge invariance of QED, that is said to be Abelian, to non Abelian (i.e. non commutative) local gauge invariance. However it seemed impossible to design a Yang Mills theory for the weak interaction that has a finite range whereas the quanta of the gauge fields in a Yang Mills theory are necessarily mass less which implies an interaction with an infinite range. As far as strong interaction is concerned, the task of describing it by means of a Yang Mills theory looked even more hopeless because of the proliferation of the family of *hadrons*, the particles that participate to all fundamental interactions including the strong one, a proliferation that seemed hard to reconcile with any sensible QFT.

A key step in the building of the standard model of particle physics was the discovery of a "sub-hadronic" level of elementarity, the level of *quarks*. These particles are spin ½ elementary constituents of hadrons, (hadronic fermions, the *baryons* are three-quark bound states, and the hadronic bosons, the *mesons* are quark-antiquark bound states) with fractional electric charges, that appeared as good candidates to be the quanta of matter fields in a gauge theory for the strong interaction, and together with the electron and other *leptons*, the quanta of a gauge theory for the weak and electromagnetic interactions.

Another major step in establishing the standard model of particle physics was the proof of the renormalisability of gauge theories, with or without spontaneous symmetry breaking, by 't Hooft, Veltman [10], Lee and Zinn-Justin [11].

For the strong interaction at the level of quarks, the non Abelian renormalisable gauge theory is Quantum Chromo Dynamics (QCD), for which the gauge group is a SU(3) group, called the "colour" group, in reference to an analogy with ordinary colours: the quarks exist in three "fundamental colours" and combine to make "white" or "colour less" hadrons. The gauge fields are the eight *gluon fields.* The coarse graining dependence of the coupling constant is translated in dependence on the resolution with which the interaction is probed, i.e. on energy. At small distance, i.e. at high energy, the QCD coupling constant is small (one says that QCD is *asymptotically free*) which leads to a reliable perturbative expansion, whereas at a distance of a typical hadronic size, the coupling diverges, which suggests that non perturbative QCD could explain the impossibility of observing free propagating quarks and gluon (the so called *confinement* of subhadronic quanta).

In the standard model, the weak and electromagnetic interactions are combined in the *electroweak* theory of Glashow [12], Salam [13] and Weinberg [14], that is, before symmetry breaking, a gauge theory, with massless *chiral* quarks and leptons as matter fields, the three massless

*intermediate vector bosons* and the photon as gauge fields. In order to obtain from this theory a reliable phenomenological model it was necessary to design the mechanism [15], known as the *Higgs mechanism*, able to generate the observed non vanishing masses of the intermediate vector bosons and of the fermions, while preserving renormalisability. This mechanism leads to the prediction of the existence of a yet non discovered particle, the *Higgs boson*. Apart from the discovery of this particle, which is its last missing link, all the predictions of the standard model of particle physics have been experimentally verified at energies up to 200 GeV. The search for the Higgs boson is the main objective of the scientific program at the LHC facility that is going to be commissioned in 2008.

## 4 General relativity and the standard model of cosmology

### 4.1 Relativity and the theoretical status of space

In the fifth appendix, untitled *Relativity and the problem of space*, added in 1952 to his book on relativity [16] for a wide public, Einstein discusses the relation between the theory of relativity and the theoretical status of space. He explains that in classical Newtonian physics, space has an independent existence, an idea that "can be expressed drastically in this way: If matter were to disappear, space and time alone would remain behind (as a kind of stage for physical happening)." In special relativity, the concept of ether, as a mechanical carrier of electromagnetic waves, is eliminated and replaced by the concept of *field*, which is one of the "most interesting events in the development of physical thought." However, in special relativity, apart from the field, there exists matter, in the form of material points, possibly carrying electric charge. This is the reason why, in special relativity, space keeps an independent existence, although, through the concept of space-time, it acquires a fourth dimension: "it appears therefore more natural to think of physical reality as a four-dimensional existence, instead of, as hitherto, the *evolution* of a three-dimensional existence."

General relativity is an extension of the theory of relativity to arbitrary change of reference frame through a detour to gravitation and with the help of the equivalence principle:
- Any arbitrary change of reference frame can *locally* be replaced by an adequate gravitational field
- Any gravitational field can *locally* be replaced by an adequate change of reference frame.

General relativity is thus a *geometric theory of gravitation*: matter and the gravitational field it generates are replaced by a non-Euclidean space-time, the metric of which is a universal field. So, Einstein could conclude his 1952 appendix by saying "On the basis of the general theory of relativity, space, as opposed to 'what fills space' has no separate existence (...) There exists no space empty of field."

### 4.2 The finite, static universe of Einstein

In terms of the metric field $g_{\mu\nu}(x)$, the Einstein's equation,

$$R_{\mu\nu}(x) - \frac{1}{2} g_{\mu\nu}(x) R = -\kappa T_{\mu\nu}(x) \tag{9}$$

relates the Ricci-Einstein curvature tensor $R_{\mu\nu}(x)$ and the scalar curvature $R$ to the energy-momentum tensor of matter $T_{\mu\nu}(x)$. In this equation, the proportionality factor $\kappa = \frac{8\pi G}{c^2}$, involving the universal constants $G$ and $c$, insures that one recovers, through this new theory of gravitation, the Newtonian theory at the non relativistic limit ($c$ going to infinity).

The study of the universe as a whole is the objective of cosmology, a domain that, until recently, belonged rather to philosophy than to science. It is not the least merit of 20[th] century physics to have provided this domain with a scientific basis through Einstein's theory of general relativity. This theoretical framework has made it possible to put together the observational data in cosmological models. A cosmological model can be obtained by modelling the material content of the universe through a specific form of the energy-momentum tensor in the right hand side of the Einstein's equation. The first cosmological model was attempted by Einstein himself. Since, at the time when he tried this model (in 1917), it was hard to imagine the universe not being static, he modified his equation that seemed not to be compatible with a static universe, by adding in the left hand side of (9) a term, fully preserving general covariance, that he called the *cosmological constant* term, which would provide a universal repulsive force (a negative pressure), preventing the universe from collapsing under the action of its own gravitation:

$$R_{\mu\nu}(x) - \frac{1}{2} g_{\mu\nu}(x) R + \Lambda g_{\mu\nu}(x) = -\kappa T_{\mu\nu}(x) \qquad (10)$$

In this model, it turns out that the universe is spatially finite but without boundary, a so called *elliptic space* with a radius $r = \sqrt{\frac{3}{\Lambda}}$. However this model has the unsatisfactory feature that the equilibrium between the action of the cosmological constant and the gravitation induced by matter is unstable, and when theoretical and observational arguments led to the conclusion that the universe is actually not static, Einstein was led to abandon his cosmological model, saying that introducing the cosmological constant was one of his biggest mistakes.

## 4.3 The standard cosmological model, from the Big Bang model to the *concordance cosmology*

4.3.1 The Hubble law

The discovery that the spectral rays of stars in distant galaxies are red shifted with a relative shift $z = \frac{\Delta\lambda}{\lambda}$ which was found to be proportional to the distance, was interpreted as an indication of a recession motion of galaxies with a velocity proportional to the distance. One thus arrived at the idea of the expansion of the universe, based on the Hubble law

$$\frac{\Delta\lambda}{\lambda} \cong H_0 \frac{L}{c} \qquad (11)$$

where $L$ is the distance of the galaxy, and $H_0$ is the Hubble constant that has the dimensional content of inverse time (the subscript 0 refers to present time, because this Hubble "constant" may have varied in the evolution of the universe). By reversing by thought the direction of time, the expansion of the universe leads to the idea that the present universe has its origin in a primordial explosion, the *Big Bang*, that is a singularity involving infinitely large density of energy and temperature that occurred some ten to twenty billions years ago.

4.3.2 The Friedman-Lemaître-Robertson-Walker metric

In the big bang model, one models the material content of the universe as a homogeneous and isotropic fluid (as a consequence of the *cosmological principle*) with an energy density $\rho$ and a pressure $p$. With such a right hand side, a solution of the Einstein's equation is given by the Friedman-Lemaître-Robertson-Walker (FLRW) metric ($c$ is set to one):

$$ds^2 = dt^2 - a^2(t)\left[\frac{dr^2}{1-kr^2} + r^2 d\theta^2 + r^2 \sin^2\theta d\phi^2\right] \qquad (12)$$

where $k$, independent on time, is equal to 1, 0 or -1, according to the spatial curvature of the universe being positive, null or negative, and where $a(t)$ is the time dependent scale factor of the universe, related to the Hubble constant by $H_0 \equiv (\dot{a}/a)_0$, the expansion rate of the universe at present.

The scale factor of the universe is governed by the two fundamental equations of "cosmodynamics":

$$\frac{\dot{a}^2 + k}{a^2} = \frac{8\pi G \rho}{3}; \quad d(\rho a^3) = -p da^3 \qquad (13)$$

The first one that depends on the spatial curvature index k relates the expansion rate of the universe to the energy density $\rho$. The second one, when coupled to the equation of state $p = p(\rho)$ which relates the pressure $p$ to the energy density, determines the evolution of the energy density as a function of the scale factor. For a spatially flat universe ($k = 0$) to occur, the energy density must have a value called the critical density $\rho_c = \frac{3H_0^2}{8\pi G}$. In phenomenological cosmology, it is convenient to parameterize the contribution of a given component of matter to energy density by its ratio to the critical density: $\Omega_i = \rho_i / \rho_c$.

4.3.3 Observational evidences for the Big Bang model

Apart from the recession of distant galaxies according to the Hubble law which is at its very origin, the Big Bang model is comforted by two other observational evidences:
- the discovery and the more and more precise measurement, in agreement with the Planck's law, of the *cosmic microwave background radiation* (CMBR), a black body radiation predicted to have been emitted when nuclei and electrons recombined into neutral atoms, and when the universe thus became transparent to photons
- The observed *relative abundance of light elements* in agreement with the one predicted by means of a model for *nucleosynthesis* in the primordial universe.

4.3.4 The inflation scenario

However the Big Bang model suffered from some drawbacks on the one hand because of the lack of precision in the observational data (in particular in the measurement of distances) and, on the other hand because of at least two theoretical problems:
- The *horizon problem* results from the principle that information cannot travel faster than light. In a universe of a finite age, this principle sets a bound, called the *particle horizon*, on the separation of two regions of space that are in causal contact. The surprising isotropy of the measured CMBR spectrum raises a horizon problem: if the universe had been dominated by matter and radiation up to the time of the emission of the CMBR, the particle horizon at that time would correspond to two degrees on the sky at present, and there is no reason why the temperatures of two regions in the CMBR map separated by more than two degree should be the same.
- The *flatness problem* belongs to the family of "fine tuning problems". It turns out that the observed density of energy is of order of the critical density corresponding to a spatially flat universe ($k = 0$); now, since any departure from the critical density grows exponentially with time, the energy density in the primordial universe should have been fine tuned with an incredible accuracy in order for the present density to be of the order of the critical density.

A possible solution to these two theoretical problems could be found in a scenario, called *inflation*, according to which there has been in the very early universe an epoch of exponential expansion of space-time such that all regions of the CMBR observed at present are in the particle horizon of each other and that the currently observable universe, that is a very small part of the whole universe, approximately appears as spatially flat. However, until very recently, this very speculative scenario was lacking observational evidence.

4.3.5 Concordance cosmology, the new cosmological standard model [17]

In recent years observational cosmology (why not calling it cosmological metrology?) has made considerable progress on the one hand in the domain of the measurement of distances of large *z* galaxies by means of the observation of supernovae of type IA, which improved the determination of the expansion rate of the universe, and on the other hand in the domain of CMBR measurement which improved the determination of the various components of the energy density of the universe. The phenomenological interpretation of the data coming from these two domains converged to what is now known as the *concordance cosmology* that is compatible with the inflation scenario, and that can be summarized as follows

   a. The spectrum of fluctuations in the CMBR is compatible with the inflation scenario
   b. The age of the universe is $13.7 \pm 0.2 \ 10^9$ years
   c. Decoupling (emission of CMBR) time is $t_{dec} = 379 \pm 8 \ 10^3$ years
   d. Our universe has $0.98 \leq \Omega_{tot} \leq 1.08$, which is compatible with spatial flatness ($k = 0$)
   e. Observation of primordial deuterium as well as CMBR observations show that the total amount of baryons (i.e. ordinary matter) in the universe contributes about $\Omega_B \cong 0.04 - 0.06$, which means that *most of the universe is non-baryonic*
   f. Observations related to large scale structure and dynamics suggest that the universe is populated by a non-luminous component of matter (dark matter, DM) that contributes about $\Omega_{DM} \cong 0.20 - 0.35$
   g. Combining this last observation with the approximate criticality of the universe (item d) one concludes that there must be one more component to energy density representing about 70% of critical density. Supernovae observation indicates that this component, called dark energy (DE) has negative pressure, for which the simplest choice is the Einstein's cosmological constant

   A comment is in order about this last item. The fact that the dark energy component of the energy density can be attributed to the effect of the cosmological constant is maybe a scientific discovery of paramount importance, namely, the discovery of a universal constant reflecting a property of the universe as a whole, its radius.

## 5 The interface between General Relativity and Quantum Field Theory

### 5.1 The tentative Grand Unification Theory (GUT) of strong and electroweak theories

From the unification of celestial and terrestrial mechanics in the Newtonian theory and the unification of electric, magnetic and optic phenomena in the Maxwell's theory, to the electroweak theory that began to unify weak and electromagnetic interactions, the quest of unity has been a very strong motivation in theoretical physics. The success obtained with the standard model of particle physics and the progresses made in observational cosmology encourage us pushing this quest further. The three non-gravitational interactions are described in terms of renormalisable gauge theories, in which the coupling "constants" measuring the intensities at the elementary level, are not constant, since they depend on energy. Now it turns out that according to renormalization group equations that govern the evolution of these couplings, they seem to converge towards the same value at energy of about $10^{15}$ GeV, which is tantamount for looking for a theory unifying the three interactions that would occur at such energy. The simplest group that contains as a sub-group the product of the three symmetry groups of the three interactions is the SU(5) group that Georgi, Quinn and Weinberg [18] considered in 1974 as the local gauge symmetry group of what has been called the Grand Unified Theory (GUT). In this theory, quarks and leptons form fivefold multiplets and there are 24 gauge bosons, the eight gluons, the four electroweak gauge bosons and twelve *leptoquarks*, i.e. gauge bosons transforming a quark into a lepton. To explain why these leptoquarks are not observable, because very massive (a mass of about $10^{15}$ GeV/$c^2$) one has to imagine a Higgs mechanism leading to two stages of symmetry breaking, one occurring at about $10^{15}$ GeV in which the leptoquarks get their mass while gluons and electroweak bosons remains massless and another one occurring at about 200 GeV in which the weak intermediate bosons become massive whereas the photon remains massless. In this theory, the exchange of leptoquarks leads to the non conservation of the baryon number, which implies that the proton could be unstable. With the mass expected for the leptoquarks, a possible lifetime of about $10^{30}$ years was estimated for the proton, which could in principle be tested experimentally.

## 5.2 The Minimal Supersymmetric Standard Model (MSSM)

The grand unified theory became doubtful because, on one hand, experimental searches did not show any evidence for the instability of the proton, on the other hand because when the evolution of the coupling "constants" related to the renormalization group equations was established with a better accuracy (that is metrology!), it appeared that they actually do not converge towards a unique value at high energy. In any case, the idea of grand unification at a very high energy scale suffers from a severe theoretical difficulty, known as the *hierarchy problem*: a Higgs mechanism involving symmetry breaking at two very different scales requires a very high fine tuning which makes it highly non credible. The attempts for curing all these drawbacks, in the same move, rely on a new symmetry, *super symmetry* (familiarly called SUSY).

Super symmetry is a new symmetry which was invented to solve certain problems of the quantization of gravity. It puts together bosons and fermions within "*supermultiplets.*" It unifies internal symmetries and space time symmetries: the square of a super symmetry transformation is the generator of space-time translations.

Since the particles of the standard model cannot be put together in supermultiplets, it should be imagined, if super symmetry there is, that it is broken in such a way that each particle of the standard model has a partner of the other statistics that has not been discovered yet. The Minimal Supersymmetric Standard Model [19] (MSSM) is a supersymmetric extension of the standard model preserving all its assets at energies up to 200 GeV, and involving the breaking of SUSY at a few hundred GeV (i.e. with partners with such masses, that could be discovered at the LHC).

This MSSM could correct all the defects of the SU(5) theory.
- With SUSY broken between 500 and 1000 GeV, one can make the three coupling "constants" to converge exactly.
- The hierarchy problem of the Higgs mechanism would be solved: in a self energy diagram, say, the boson loops and fermions loops have contributions of different signs which tend to compensate each other, in such a way that the quadratic divergences at the origin of the hierarchy problem disappear
- In a gauge theory with SUSY, the Higgs fields and the potential of the Higgs mechanism are constrained, whereas in a theory without SUSY, they are completely ad hoc.
- The lightest neutral supersymmetric partners (called "neutralinos") would be stable particles, interacting only very slightly: they are excellent candidates for solving the cosmological problem of dark matter.
- The superstring theory currently considered as the best candidate for unifying all fundamental interactions, including gravitation inevitably implies supersymmetry.

# 6 Quantum statistics and modern quantum physics

## 6.1 The modern interpretation of quantum physics: emergence of a quasi-classical realm

Quantum Statistics is this part of quantum physics that does not deal with "pure states" but that deals with mixed states, i.e. with partially known states. Since partial knowledge is due to coarse graining, one can say that quantum statistics is nothing but *coarse grained quantum physics*. The fact that there exists no reliable phenomenological description of physical reality without some coarse graining reflects the fundamental limitation principle related to the universal constants $h$ and $k$. In this respect it is interesting to notice that in the path integral formulation of QFT, the Planck's constant is not actually interpreted as a limitation, it just provides the unit of action necessary to define the integrand of the functional path integral. The fact that the limitation aspect of quantum physics is relegated to the domain of phenomenology, experiment and observation is maybe the reason why the PI formulation of QFT is so well suited for free theoretical investigations and explorations.

Quantum Statistics is fully integrated in the orthodox (Copenhagen) interpretation of quantum physics that has not suffered any contradiction with experiment in about eighty years. One can say that, quantum physics supplies, through suitable coarse graining, satisfactory descriptions for all phenomena observable in the world at our scales [20] from the atomic to the galactic scales. However, for our purpose of establishing a link between phenomenology and the fundamental theories underlying the standard models of particle physics and cosmology, the orthodox interpretation may look insufficient because it postulates the existence of a classical world together with a quantum world and because it seems to require an observer outside the system making measurements. If one wants an interpretation of quantum physics that explains the emergence of the classical world and that authorizes quantum effects to occur in the primordial universe (when there was no "observer" whatsoever), one has to generalize the interpretation of quantum physics.

According to the interpretation of quantum physics in terms of decoherent alternative histories that performs this generalization [21], the object of quantum physics is to supply probabilities for sets of alternative histories of the universe, by means of a path integral functional, called the *decoherence functional*. In order to be able to attribute probabilities to alternative histories, one has to perform the path integral with a graining that is sufficiently coarse so that interferences that might prevent ascribing additive probabilities to independent events actually cancel: this is what is meant by the term *decoherence*. Coarse graining beyond the one necessary for decoherence is needed to achieve approximate deterministic predictability of classical physics. A set of coarse-grained alternative histories is called a *quasi-classical realm*. The coarse graining leading to such a quasi-classical realm is connected with the coarse graining leading to the familiar entropy of thermodynamics. The probabilities attributed to alternative histories of the universe are *a priori* probabilities, and as Gell-Mann notes [22] in his article untitled *What is Complexity?* "These are a priori probabilities rather than statistical ones, unless we engage in the exercise of treating the universe as one of a huge set of alternative universes, forming a 'multiverse.' Of course, even within a single universe cases arise of reproducible events (such as physics experiments), and for those events the a priori probabilities of the quantum mechanics of the universe yield conventional statistical probabilities".

In any case, the modern interpretation of quantum physics aims at explaining how the quasi-classical realm, in which we live and possibly perform metrology experiments, emerges from the primordial universe describable in terms of the fundamental theories (QFT and GR).

## 6.2 The interface between Quantum Field Theory and Quantum Statistics

6.2.1 Renormalisable QFT and critical phenomena

The functional integral over locally coupled fields $\phi$ in QFT in 3+1 dimensions can be translated into the Boltzmann-Gibbs classical probability distribution for a configuration $S$ of a four dimensional system of spins coupled to near neighbours by means of a strict "dictionary"

$$\mathcal{Z}_{QFT} = \int \mathcal{D}\phi \exp\left(\frac{i}{\hbar}\int dt dx_1 dx_2 dx_3 \mathcal{L}(\phi)\right) \Leftrightarrow \mathcal{Z}_{Classical} = \int \mathcal{D}S \exp\left(-\frac{1}{kT}\int dx_1 dx_2 dx_3 dx_4 \mathcal{H}(S)\right)$$

$$\hbar \Leftrightarrow k$$

$$idt \Leftrightarrow \frac{dx_4}{T} \qquad (14)$$

$$\mathcal{L}(\phi) \Leftrightarrow \mathcal{H}(S)$$

This dictionary can be used to translate quantum field theoretic problems into classical statistical problems and vice-versa: for instance, to the perturbative expansion in QFT corresponds, in classical statistical physics, the calculation of corrections to the *mean field approximation*. Now it turns out that in the case of second order phase transitions (leading to *critical phenomena*), the mean field approximation breaks down because the correlation length goes to infinity and the system undergoes *self similar fluctuations* at all scales. K. Wilson has shown that the method used to solve this difficulty in statistical physics corresponds, in the above mentioned correspondence (14), to the renormalization procedure in a renormalisable theory, which extends the above mentioned dictionary to the very fruitful new item: *critical phenomena occurring at the second order phase transition correspond to the results of a renormalisable quantum field theory.*

In statistical physics, the difficulty encountered in second phase order transitions is comparable with the divergences which, in QFT, prevent from applying perturbative methods: the presence of fluctuations at all scales indicates that the behavior of the system seems to depend on the details of the interaction at the microscopic level, just as the regularization in QFT seems to make physics to depend on an arbitrary high energy cut-off parameters. The recursion method, due to Kadanoff [23], used to average, scale after scale, the self similar fluctuations affecting the system at a second order phase transition consists in looking for a *fixed point* in an ensemble of systems corresponding to varying coarse graining, related to one another by a transformation, called by Wilson [24] a *renormalization group transformation*. Near such a fixed point, the critical phenomenon somehow "forgets" the lattice spacing, just as in a renormalisable QFT, one can without disadvantage let the cut-off parameter go to infinity (see paragraph 3.3.3).

Just as, in a renormalisable QFT, renormalized Green's functions and renormalized parameters depend on the coarse graining, the dynamics near the fixed point of the systems related by renormalization group transformations depend on the coarse graining. But, in the same way as in critical phenomena, fixed point equation expresses self similarity, i.e. what is invariant under a change of scale, in a renormalisable QFT, the renormalization group equation expresses what, in the dynamics is invariant under a change in coarse graining.

The correspondence we have just described is at the origin of a remarkable interdisciplinary synthesis that occurred in the 70's, since one was able to use, with great success, the same theoretical tools in two domains of physics which hitherto seemed completely disconnected, the physics of critical phenomena on one hand and Quantum Chromo Dynamics, the physics of strong interaction on the other hand. In particular, this correspondence allowed designing some computer simulations of QCD, the so-called "lattice QCD" providing some insight on the non-perturbative regime of this quantum field theory.

6.2.2 Emergence, universality and effective theories

The epistemological lesson of this success is that it helps in understanding what the meaning of the concept of emergence is. In a second order phase transition, the behaviour of the system showing correlations with a range much larger than the lattice spacing, with giant fluctuations, is what one typically calls an *emergent phenomenon*. Since systems which are very different at microscopic scales like a spin lattice and a renormalisable QFT may show similar emergent behaviours, one can say that what is emergent exhibits some sort of *universality*. Actually, of two such systems that share the same renormalization group equation, one says that they belong to the same *universality class*. Another important feature of emergence is that universal emergent behaviour depends only on a few *relevant degrees of freedoms or parameters*, whereas all other degrees of freedom or parameters can be considered as *marginal*. It is thus legitimate to say that two systems, one as simple as a spin lattice suitable for a computer simulation, and another one as sophisticated as QCD, which belong to the same universality class, differ only by marginal degrees of freedom or parameters. This remark is at the heart of the *methodology of effective theories* [25] that provides the link between the theoretical framework and phenomenological modelling. The basic idea of effective theories is that if there are parameters that are very large or very small with respect to the relevant physical quantities (of the same dimension), one can obtain a much simpler physical approximate description by putting to zero the very small parameters and to infinity the very large ones. Actually the standard model of particle physics is not its "fundamental theory"; it is but a set of effective theories adequate for the description of physics up to energy of a few hundred GeV, all the "effects" that are used in quantum metrology rely on effective theories adequate for the decription of physics in our quasi-classical realm.

**6.3 The interface between General Relativity and Quantum Statistics**

### 6.3.1 The concept of horizon in Quantum Statistics and in General Relativity

In quantum statistics, the concept of *horizon* is relevant in the interpretation of the fundamental limitations of human knowledge implied by the Planck's and Boltzmann's constants: these limitations are not to be considered as insuperable obstacles but rather as *informational horizons*, namely some boundaries defined in an abstract space, beyond which lie some inaccessible information. The fact of assuming the existence of an informational horizon does not mean that one neglects or forgets the information lying beyond it. The methodology that allows keeping track of this missing information is based on functional integration: to evaluate the probabilities of the values of the dynamical variables bearing the accessible information (the followed variables) one *integrates out* the dynamical variables bearing the inaccessible information (the non-followed variables). Basically, this is the meaning of the methodology of effective theories sketched in the preceding paragraph.

In general relativity gravitation appears as an interaction (actually the only one) capable of so much curving space-time that it leads to the formation of a *geometric horizon*, namely a light-like surface acting as a "one-way membrane", a two-dimensional informational horizon hiding information lying beyond it. Because of the expansion of universe, such a horizon obviously exists in cosmology: it is defined by the distance at which lie galaxies whose light is infinitely red-shifted. A theoretical laboratory to explore the physics of such horizons is the physics of black holes. The horizon of a black hole, called its *event horizon*, is the surface surrounding it of infinite gravitational red-shift, beyond which any matter (and thus any information), trapped by the black hole escapes from perception. At the horizon, the gravitational field is so intense that it may induce in matter certain quantum effects such as the production of particle antiparticle pairs, which have to be dealt with. Since, in quantum statistics, missing information is equivalent to entropy, it is natural, in this framework, to attribute entropy to such a horizon. Bekenstein and Hawking have shown that the entropy corresponding to the information trapped inside a black hole is $k$ times one quarter of the area of the horizon in Plank's units.

$$S = k \frac{1}{4} \frac{A_{\text{Horizon}}}{A_P} = \frac{kc^3}{\hbar G} \frac{1}{4} A_{\text{Horizon}} \qquad (15)$$

### 6.3.2 A promising new concept: Holography

Since entropy is hidden information, (15) can be interpreted as a bound, the so called Bekenstein's bound [26], on the information that can be packed in the volume of a black hole. The finiteness of this bound leads to the conjecture that the world has to be discrete, for, would it be continuous, and then every volume of space would contain an infinite amount of information. The fact that the Planck's units appear in (15) suggests that the discrete structure of space-time that it involves is related to quantum gravity. This reasoning can be generalized to any horizon (and even to any surface) in terms of what one calls the *holographic principle* [27], that Lee Smolin [28] states as follows: "Consider any physical system, made of anything at all, let us call it The Thing. We require only that The Thing can be enclosed within a finite boundary, which we shall call The Screen. (…) Since the observer is restricted to examining The Thing by making observations through The Screen, all what is observed could be accounted for if one imagined that, instead of The Thing, there was some physical system defined on the screen itself. This system would be described by a theory which involved only The Screen. (…) If the screen theory were suitably chosen, the laws of physics inside the screen could equally well be represented by the response of The Screen to the observer."

It is interesting to note that in the correspondence (14) between quantum and classical statistical physics, one has something that is similar to the holographic principle: since, in this correspondence, imaginary time corresponds to a fourth Euclidean dimension, one can say that somehow quantization adds an extra space dimension to classical physics: quantum physics in a three-dimensional space is equivalent to classical physics in a four-dimensional space.

# 7 Conclusion: universal constants, fundamental metrology and scientific revolutions

## 7.1 The Copernican revolution

Let us conclude this chapter by reviewing the role of universal constants and fundamental metrology in the occurrence and accomplishment of scientific revolutions. Consider first the Copernican scientific revolution. It marks the outset of modern science with the Newton's theory of universal gravitation taking into account the constant $G$ and unifying terrestrial and celestial mechanics. We then have the accomplishment of this scientific revolution at the turn of the $20^{th}$ century by means of the first standard models consisting on
- The Newton's theory of universal gravitation
- the unification of rational mechanics and the atomistic conception through the kinetic theory of matter and statistical thermodynamics that take into account the constant $k$
- the unification of electricity, magnetism and optics through the Faraday-Maxwell-Hertz theory that takes into account the constant $c$.

The Kepler's laws, the accurate measurements of the velocity of light, the discovery of Neptune, hypothesized to explain the anomalies in the motion of Uranus, the thirteen independent measurements by Jean Perrin of the Avogadro number that proved in an irrefutable way the existence of atoms, are all elements belonging to the realm of metrology that played an essential role in the establishment and the accomplishment of the Copernican revolution/

## 7.2 Universal constants and the crisis of physics at the beginning of the $20^{th}$ century

The beginning of the last century was marked by a severe conceptual crisis related to the discovery or re-discovery of universal constants:
- The Boltzmann's constant $k$ is connected to the critical problem of inobservability, identity and stability of the atoms, the elementary constituents of matter
- The velocity of light $c$ is connected to inconsistency of the model of ether
- The Planck's constant $h$ is connected to the objectivity, causality and determinism crisis.

## 7.3 The $20^{th}$ century scientific revolution

This is the revolution that we have described in this chapter and that consists on the re-foundation of the theoretical framework by means of quantum field theory, general relativity and quantum statistics that take into account pairs of the four universal constants. The accomplishment of this revolution consists on the validation of the standard models that gives an acceptable agreement with all experimental and observational data by means of a fixed and finite number of free parameters, the dimensionless universal constants, the determination of which relies on fundamental metrology. The stake of the LHC program the main objective of which is the search for the *very last missing link of this revolution* is particularly high. This revolution is characterized by its complete openness to new experimental or observational discoveries (in this respect the LHC, Planck, VIRGO and LIGO programs appear extremely promising) and to new theoretical ideas or principles at the interfaces of the three theories of the tripod (supersymmetry, superstring, loop quantum gravity, non commutative geometry holographic principle, membranes, new dimensions of space…).

## 7.4 Is the cosmological constant the fifth universal constant signalling a new scientific revolution at the horizon?

What we have said about holography puts on the foreground a striking analogy with the situation at the beginning of the $20^{th}$ century. At that time it was the attempt to understand the thermodynamic properties of the electromagnetic radiation that induced progress in the understanding of atomic physics, and that led to the quantum physics revolution. Today it seems that it is the attempt to understand the meaning of the entropy of a horizon (of a black hole or the cosmos with a non vanishing cosmological constant) that may open, through the holographic principle, a thermodynamic route [29] towards quantum gravity. On the other hand, the crisis induced by the cosmological constant whose QFT and GR interpretations disagree by more than 120 orders of magnitudes is at least as severe as the crisis that aroused at the beginning of the 20$^{th}$ century in relation with quantum physics.

What we can say for sure is that the energy density, space and time scales of the domain of quantum gravity are so extreme that one will never be able to explore it by direct experiments. One will necessarily rely on theoretical modelling leading, in observable quantities to deviations with respect to the standard models ignoring quantum gravity. Very accurate metrology to detect and ascertain such deviations will be the main tool, maybe the only one, available to prove or disprove any future theory of quantum gravity. Fundamental metrology has thus a very promising future.